**Physicochimie des interfaces chargées :**

**modélisation multi-échelle et applications pour l'énergie**

Nous présentons l'intérêt d'une démarche de modélisation multi-échelle pour la compréhension des systèmes présentant des interfaces chargées. D'une part, il s'agit de simuler un même système complexe à différentes échelles, en fonction des échelles d'espace et de temps pertinentes pour répondre à une question de physico-chimie donnée. D'autre part, il convient de faire le lien entre les différents niveaux de description, par exemple dans le cadre d'une approche ascendante "bottom-up". Nous illustrons plus particulièrement ici le cas des matériaux poreux chargés que sont les argiles, en abordant des problèmes de physico-chimie dans les contextes du stockage des déchets nucléaires en couche géologique profonde et du stockage géologique du $CO_2$.

**Mots-clés**: Modélisation multi-échelle, coarse-graining, interface solide-liquide, ions, argiles

**Physical chemistry of charged interfaces :**

**Multiscale modelling and applications to Energy**

We present the advantages of a multiscale modelling strategy for the understanding of systems with charged interfaces. On the one hand, one can simulate a complex system at different levels, depending on the relevant length and time scales for a given physical chemistry problem. On the other hand, one should make the link between the various levels of description, e.g. following a bottom-up approach. We illustrate here the case of charged porous materials, in particular clay minerals, by discussing physical chemistry issues that arise in the context of geological disposal of nuclear wastes and $CO_2$ sequestration.

**Auteur**

Benjamin Rotenberg

Chargé de recherche CNRS au laboratoire Physicochimie des Electrolytes et Nanosystèmes Interfaciaux (PHENIX, UM3 8234) à l'UPMC (Université Pierre et Marie Curie).

**Introduction**

La plupart des matériaux d'intérêt industriel et environnemental, qu'ils soient naturels ou manufacturés, présentent des hétérogénéités depuis l'échelle atomique et moléculaire jusqu'à l'échelle macroscopique, ainsi qu'aux échelles intermédiaires (nm-µm notamment). Lorsqu'ils sont mis en contact avec un liquide, dans la nature ou pour une application donnée, leur surface est souvent chargée du fait de leur composition chimique et de la dissociation de sites de surfaces. Nous discuterons plus particulièrement ici le cas des argiles, matériau naturel poreux et chargé, et nous aborderons des questions de physico-chimie en lien avec leur rôle dans le stockage des déchets nucléaires en couche géologique profonde [1] et le stockage géologique du $CO_2$ (FIGURE 1)

Du point de vue du physico-chimiste, la difficulté à appréhender ces systèmes vient de la complexité des interactions mises en jeu, depuis l'échelle moléculaire jusqu'à celle du matériau : réactivité des sites de surface, solvatation des ions dans le volume et à l'interface, mouillabilité, hydrodynamique en milieu nano-confiné, effets électrocinétiques, etc. Ainsi, le transport des ions aux interfaces chargées et leur sorption (fixation à la surface) met en jeu des échelles de longueur et de temps couvrant plusieurs ordres de grandeur, de l'angström (1 Å = $10^{-10}$ m) et la picoseconde (1 ps = $10^{-12}$ s) qui sont typiques des processus chimiques moléculaires, à plusieurs nanomètres (soit $10^{-9}$-$10^{-7}$ m) et la microseconde (1 µs = $10^{-6}$ s) où se manifestent les effets interfaciaux (comparés au comportement dans le volume). Dans les matériaux poreux, les hétérogénéités caractéristiques de ces interfaces sont de plus couplées aux hétérogénéités géométriques à plus grande échelle (> $10^{-8}$ m).

Du point de vue de la modélisation, le défi représenté par les multiples échelles de temps et d'espace ne peut être relevé que par une approche de modélisation multi-échelles. Ce concept, dont on entend souvent parler ces dernières années, couvre en réalité plusieurs aspects. Il d'agit d'une part de modéliser un même système à plusieurs échelles pour répondre à des questions scientifiques (ici de physico-chimie) et prendre en compte différents types de complexité. D'autre part, un aspect moins souvent abordé porte sur le lien entre les différents niveaux de description pour obtenir des modèles "aussi simples que possible, mais pas plus" et pour assurer une cohérence entre les différentes façons de décrire le même système (FIGURE 2). Outre le cas des argiles présenté ici, on peut citer par exemple la mise en oeuvre de cette stratégie pour les matériaux cimentaires [2], les métaux et alliages métalliques [3] ou encore la matière molle [4].

Nous ne présenterons pas ici l'ensemble des méthodes de simulation disponibles, ni l'ensemble des stratégies permettant de faire le lien entre les différentes échelles. Sur ce dernier aspect, il faut en particulier noter les approches dites "hybrides", c'est-à-dire couplant explicitement au sein d'une même simulation des sous-systèmes décrits à des niveaux différents, telles que les simulations mixtes quantique/classique, souvent utilisée pour décrire les sites actifs de biomolécules et dont la mise en oeuvre par Karplus, Levitt et Warshel a été récompensée par le Prix Nobel 2013 de Chimie [5,6]. Cette stratégie présente l'avantage de limiter le temps de calcul en concentrant les efforts sur la partie du système qui nécessite la description la plus fine. La difficulté principale est alors de décrire correctement le couplage entre les différents sous-systèmes. Les échelles de temps accessibles restent par ailleurs limitées par le processus le plus rapide, qui fixe la durée totale du phénomène que l'on peut simuler.

Après une brève description de la modélisation et la simulation à différentes échelles, nous présentons plusieurs exemples illustrant l'utilisation des niveaux de description les plus fins pour construire des modèles plus simples. Puis, nous montrerons l'apport de différents niveaux de modélisation pour la compréhension et la simulation des phénomènes électrocinétiques, c'est-à-dire le couplage entre effets électriques et hydrodynamiques qui se manifestent au voisinage des interfaces chargées. Nous illustrerons enfin comment la simulation permet de résoudre différents problèmes de physico-chimie des argiles en étudiant ces systèmes à l'échelle moléculaire.

**I. Modélisation multi-échelles**

*Simuler à différentes échelles*

Dans la seconde moitié du XX$^e$ siècle, la simulation, sous ses différentes formes, est devenue un outil incontournable pour la compréhension de la matière, complémentaire de l'expérimentation et de la théorie. Au cours d'une simulation moléculaire, on cherche à décrire les propriétés d'un système physique en le soumettant à une "expérience numérique" : on le place dans des conditions données (par exemple dans l'ensemble canonique, c'est-à-dire à nombre de particules, volume et température fixés), on effectue des mesures (comme la position et la vitesse de chaque atome) et on en déduit des propriétés observables (telles que l'énergie du système, des profils de densités, etc). Une différence fondamentale avec les expériences est que le système évolue selon les prescriptions que nous lui imposons, à savoir (a) un algorithme permettant de calculer, à partir des positions et des vitesses des atomes à un instant donné, ces quantités à l'instant ultérieur, et (b) une description des interactions entre atomes permettant le calcul des forces exercées sur chaque atome à partir de toutes les positions. Il s'agit donc de choisir, en fonction du problème posé, le meilleur compromis entre, d'une part, la précision avec laquelle on peut représenter le système réel et les interactions entre ses constituants, et, d'autre part, la nécessité de le décrire sur les échelles d'espaces et de temps qui correspondent au phénomène considéré,

A l'échelle atomique, les méthodes les plus développées sont la dynamique moléculaire et la méthode de Monte-Carlo [7] où chaque atome est explicitement représenté. Dans les deux cas, on distingue également les simulations *ab initio* et les simulations moléculaires classiques (Encadré 1). A l'échelle mésoscopique, il n'est pas possible (ni utile) de suivre la trajectoire individuelle de chaque atome ou molécule. On peut alors adopter une représentation simplifiée du système, en remplaçant des groupes d'atomes par des "gros grains" (*coarse-grains*) dans une simulation Monte Carlo, de dynamique moléculaire ou brownienne. On peut même simplifier encore plus la description, en termes de profils de densité et de vitesse de chaque espèce, comme on le fait par exemple lorsque l'on résout l'équation de Stokes pour décrire un écoulement ou l'équation de Poisson-Boltzmann pour décrire les distributions ioniques au voisinage d'une surface chargée. Un niveau de description intermédiaire, reposant sur la théorie cinétique (introduite par Ludwig Boltzmann pour l'étude des gaz), permet de garder des informations statistiques sur les trajectoires de chaque molécule tout en garantissant l'émergence du comportement macroscopique. Ce type de représentation se prête également bien à la simulation numérique, en particulier par les méthodes dites "sur réseau" (Encadré 2).

*Passer d'un niveau de description à l'autre*

Avant d'illustrer l'utilité de ces différentes méthodes pour répondre à diverses questions de physico-chimie, voyons d'abord l'intérêt d'inscrire l'ensemble de ces approches dans une démarche multi-échelle. On peut en effet utiliser les descriptions les plus fines (donc les plus limitées en termes d'échelles d'espace et de temps) pour établir ensuite de bonnes représentations aux échelles supérieures. Cette démarche "bottom-up" de passage de paramètres, appelée dans ce contexte *up-scaling* ("passage à l'échelle") ou *coarse-graining* (qui pourrait se traduire par "gros-grainisation", qui renvoie à l'idée de rendre la représentation plus grossière), est en fait couramment adoptée, sans forcément en avoir conscience. Lorsqu'une propriété nécessaire à la simulation à l'échelle macroscopique (par exemple la densité ou la viscosité d'un fluide pour un calcul d'écoulement dans un problème de génie des procédés) n'est pas disponible expérimentalement, il est possible de la calculer par une simulation moléculaire - à condition bien sûr que l'on dispose d'un bon modèle, validé par ailleurs, de ce système. Mais ce n'est pas le seul intérêt, car il est en général possible de tirer plus d'informations que de simples paramètres, en testant par exemple la validité de certaines hypothèses.

Un premier exemple de cette approche concerne le passage d'une description quantique, prenant en compte explicitement les électrons, à des modèles d'atomes et molécules interagissant de façon classique, l'effet des électrons étant alors pris en compte de façon implicite. En effet, la pertinence des résultats des simulations moléculaires dépend beaucoup de la qualité des modèles utilisés pour représenter le système étudié: le développement de champs de force réalistes est ainsi un champ de recherche très actif. Plusieurs voies sont possibles pour y parvenir, l'une d'entre elles consistant à exploiter des résultats de simulation *ab initio* ou de chimie quantique [8]. Nous avons donc proposé une méthode de construction systématique de champs de force classiques à partir de simulations au niveau DFT (théorie de la fonctionnelle de la densité), en utilisant le concept d'orbitales de Wannier maximalement localisées. Il était déjà possible d'obtenir de bons champs de forces polarisables par ajustement numérique des paramètres sur les forces et les dipôles atomiques calculés au niveau DFT [9]. Nous avons montré qu'en faisant certaines hypothèses, on peut, grâce aux orbitales de Wannier, déterminer séparément la quasi-totalité de ces paramètres, ce qui réduit le risque de compensation d'erreurs. Si cette approche complémentaire à l'ajustement numérique ne permet pas toujours d'obtenir des descriptions aussi fines des interactions, elle offre une meilleure compréhension de l'origine des différents termes et illustre la pertinence des orbitales localisées pour décrire les propriétés individuelles des atomes en phase condensée [10]. Ce travail s'est prolongé pour les ions dans l'eau [11] et les argiles. Pour les ions dans l'eau ($Li^+$, $Na^+$, $K^+$, $Rb^+$, $Cs^+$, $Mg^{2+}$, $Ca^{2+}$, $Sr^{2+}$ et $Cl^-$), les champs de force ont été validés par comparaison avec des grandeurs expérimentales portant sur la structure, la thermodynamique (enthalpie libre d'hydratation) et la dynamique (coefficient de diffusion de ces ions). A titre d'exemple, le signal EXAFS simulé pour l'ion $Ca^{2+}$ hydraté est comparé directement avec le signal expérimental sur la FIGURE 3.

Il est également possible de proposer une démarche analogue de simplification des interactions pour représenter une molécule complexe par un petit nombre de sites correspondants à plusieurs atomes, souvent groupés en "gros grains" en suivant l'intuition chimique. Dans un contexte différent de celui développé ici, nous avons ainsi travaillé à la construction de modèles à gros grains pour les liquides ioniques, constitués exclusivement d'ions, sans aucun solvant. Pour les applications visées, il s'agit de cations organiques tels que le 1-butyl-3-méthylimidazolium ($BMI^+$) ou le 1-éthyl-3-méthylimidazolium ($EMI^+$) et d'anions tels que l'hexafluorophosphate ($PF_6^-$) ou le tétrafluoroborate ($BF_4^-$). Un tel modèle à gros grains avait donné de bons résultats pour reproduire tant les propriétés structurales et dynamiques dans le volume [12] que la tension interfaciale ou la structure à l'interface avec des feuillets de graphite [13]. Le champ de force a été complété afin de modéliser également les ions $EMI^+$ et $BF_4^-$ pour pouvoir simuler les mélanges de ces différents ions [14]. Le modèle gros-grains obtenu reproduit bien, pour les liquides ioniques BMI-BF4 et EMI-BF4, les résultats expérimentaux ou de simulations tout-atome sur un ensemble de propriétés structurales et dynamiques au cœur des liquides : densité, fonctions de distributions radiales, coefficients de diffusion et viscosité entre 298 et 500 K (FIGURE 4). Il reproduit également la tension de surface liquide-vapeur à 400 K, ainsi que les propriétés capacitives et structurales de ces liquides entre des électrodes de graphite. Ces modèles nous ont permis d'élucider ensuite le mécanisme microscopique à l'origine des performances des supercondensateurs dont les électrodes sont constituées de carbone nanoporeux [15].

Un dernier exemple de passage d'un niveau de description plus fin vers un niveau plus approché, où le système est cette fois représenté de façon continue, concerne les écoulements dans des pores de taille nanométrique. Pour ces situations de confinement extrêmes, la validité d'une description continue n'est pas garantie. Dans le cas d'une surface d'argile de type montmorillonite, des simulations de dynamique moléculaire hors d'équilibre nous ont ainsi permis de montrer que pour des tailles de pore supérieures à 4 nm, l'hydrodynamique du fluide au-delà d'environ 1,5 nm des surfaces est bien décrite par l'équation de Navier-Stokes, avec une viscosité égale à celle du liquide non-confiné [16]. Il est cependant nécessaire de prendre en compte le glissement à l'interface solide/liquide, avec une longueur de glissement déterminée par les simulations moléculaires (environ 2 Å pour l'argile montmorillonite sodique).

## II. Simulation multi-échelles des phénomènes électrocinétiques

Au voisinage d'une surface chargée, le fluide est également chargé par la présence des contre-ions compensateurs, sur une distance caractéristique, appelée longueur de Debye, résultant de la compétition entre l'attraction électrostatique de ces ions pour la surface et le gain entropique à explorer l'ensemble du liquide. La charge du fluide conduit à un ensemble de phénomènes hydrodynamiques et électriques couplés, appelés phénomènes électrocinétiques. Par exemple, l'application d'un champ électrique tangentiel à la surface accélère le fluide dans cette région chargée, appelée double-couche, puis l'ensemble du fluide est mis en mouvement par entraînement visqueux (diffusion de la quantité de mouvement), y compris dans la région où le fluide est électriquement neutre. Ces phénomènes ont des conséquences importantes et des applications dans des domaines aussi variés que la microfluidique (pompage électrocinétique), les systèmes colloïdaux (mobilité électrophorétique), la biologie (canaux ioniques) ou les milieux poreux comme les roches (effet séismo-électrique). On peut également observer ce type de phénomènes à une interface entre deux liquides.

Les échelles d'espace pertinentes vont ainsi de l'échelle moléculaire, qui contrôle l'éventuel glissement du fluide à la surface ainsi que la solvatation des ions en contact avec la surface (complexes de sphère interne ou externe), à l'échelle mésoscopique de la double-couche (typiquement quelques nm dans l'eau en fonction de la concentration en sel, mais jusqu'à plusieurs micromètres dans des solvants organiques moins polaires) et des pores (dont la taille et la géométrie sont spécifiques au matériau, et pas forcément bien connues expérimentalement), et l'échelle macroscopique de l'échantillon. A ces différentes échelles correspondent bien sûr des temps caractéristiques variant sur plusieurs ordres de grandeur, Des stratégies variées ont été proposées pour la simulation des ces phénomènes [17].

*Echelle moléculaire : effets spécifiques et glissement*

Les simulations moléculaires évoquées ci-dessus ont ainsi permis de caractériser la structure et la dynamique de l'eau et des ions sur les surfaces basales de montmorillonite. Un exemple d'effet spécifique, c'est-à-dire dépendant de la nature chimique des ions, porte sur la structure des complexes formés à la surface, majoritairement de type sphère externe pour les contre-ions $Na^+$ et de type sphère interne pour les contre-ions $Cs^+$. La longueur de glissement mesurée à la surface lors de l'électro-osmose est la même que celle déterminée pour un écoulement sous l'effet d'un gradient de pression [18].

*Echelle mésoscopique : couplages électrocinétiques*

Si les simulations moléculaires permettent de préciser ce qui se passe dans les premières couches de solvant et de comprendre les effets spécifiques des ions, une approche mésoscopique est nécessaire pour simuler ces phénomènes pour des échelles supérieures à 10 nm et 10 ns. Cette description consiste du point de vue thermodynamique en une fonctionnelle d'énergie libre (DFT classique) et du point de vue dynamique en un bilan de quantité de mouvement (équation de Navier-Stokes) couplée à une évolution de la composition (de type DFT dépendante du temps), le tout étant implémenté par un algorithme de type "Lattice-Boltzmann" (ENCADRE 2). Nous avons étendu le champ d'application de cette méthode aux milieux poreux chargés (solide/liquide) et aux interfaces liquide-liquide (eau/organique), où l'on peut observer une séparation de charge si anions et cations n'ont pas la même affinité pour chacune des deux phases [19]. Dans le cas des milieux poreux, nous avons ainsi pu reproduire certaines observations expérimentales concernant l'évolution du coefficient de diffusion macroscopique de traceurs radioactifs ($^{22}Na^+$, $^{36}Cl^-$ et eau tritiée HTO) à travers l'argile, en fonction de leur charge et de la concentration en sel dans le milieu, et en proposer une interprétation microscopique en termes de chemins suivis au sein de la porosité du matériau. Plus récemment [20], nous avons introduit dans ce type de simulations les réactions de sorption/désorption à l'interface solide/liquide, qui sont nécessaires à la description complète du transport d'ions dans les argiles à cette échelle, et qui pourront également trouver des applications dans d'autres contextes comme la chimie analytique et séparative [21].

*De l'échelle du pore à l'échelle de l'échantillon*

Cette approche de simulation mésoscopique se révèle également utile pour faire le lien avec le transport dans les milieux poreux à l'échelle macroscopique. Il est très difficile de caractériser expérimentalement la microstructure à l'échelle de 10 nm - 1 μm des argiles compactées. Nous avons déterminé les paramètres de transport des ions à l'échelle du pore (en fonction de la taille du pore, la charge des surfaces, la force ionique de la solution) pour les introduire dans un modèle de type "réseau de pores" à plus grande échelle (Pore Network Model, ou PNM). L'objectif est de parvenir à une distribution de taille et connectivité des pores qui rende compte des observations expérimentales à l'échelle de l'échantillon. Nous obtenons ainsi un "échantillon numérique" de plusieurs μm de côté sur lequel nous pourrons ensuite étudier l'évolution des propriétés de transport dans différentes conditions pertinentes pour le stockage des déchets radioactifs, comme la désaturation ou le colmatage de certains pores en fonction de l'humidité relative ou de la concentration en sel. Les simulations de type Lattice Boltzmann permettent à la fois de préciser le domaine de validité de théories analytiques simplifiées, et de fournir les résultats numériques lorsque l'on en sort [22]. Ces modèles pourraient être utiles pour d'autres applications mettant en jeu des membranes chargées, par exemple pour la désalinisation de l'eau de mer.

**III. Physicochimie des argiles : simulations à l'échelle moléculaire**

*Acidité des sites de bordure de feuillet*

Les feuillets d'argile ne sont pas infinis et les surfaces latérales des particules présentent des propriétés différentes des surfaces basales (FIGURE 1) car elles résultent de la rupture de liaisons chimiques (Al-O et Si-O principalement). Par réaction des atomes insaturés à la surface avec des molécules d'eau, on obtient alors différents groupements chimiques selon les faces cristallographiques considérées. Par exemple, pour la face (010) de la pyrophyllite et de la montmorilllonite, on trouve des sites aluminols AlOH et $AlOH_2$ ainsi que des silanols SiOH (FIGURE 5). Ces sites sont très importants, car ils peuvent se déprotoner, donc se charger négativement et constituer des sites de sorption pour les cations. Dès lors, il est nécessaire de prévoir l'état de protonation des différents sites de surface en fonction des conditions d'acidité du milieu, c'est-à-dire du pH de la solution. Expérimentalement, on n'a accès par titration qu'au comportement global, dont il faut déduire le $pK_a$ des différents sites de surface sans que l'on puisse sonder directement l'acidité de chacun d'entre eux. Pour prévoir la sorption des ions, nous avons évalué ces $pK_a$ par une méthode consistant à calculer, par dynamique moléculaire *ab initio*, l'énergie libre de la réaction de transfert de proton entre le site de surface et une molécule d'eau loin de la surface [23]. Le site SiOH est légèrement plus acide que $AlOH_2$, avec des $pK_a$ proches de 7 pour les deux sites, tandis que AlOH ne se déprotone pas en phase aqueuse ($pK_a \approx 22$). La contribution de l'énergie de réorganisation à l'énergie libre de la réaction est importante (FIGURE 5). La forte stabilisation du site $SiO^-$ après déprotonation s'explique ainsi par l'arrivée de deux molécules d'eau donnant des liaisons H au site ainsi formé.

*Contact avec un réservoir de $CO_2$*

Dans le cadre du stockage géologique de ce gaz à effet de serre, nous avons étudié l'interaction d'un réservoir de $CO_2$ avec une couverture argileuse (FIGURE 6). Le calcul de l'énergie libre de gonflement par simulations Monte-Carlo dans l'ensemble grand-canonique montre que l'on ne doit pas s'attendre à un retrait de l'argile en présence de $CO_2$ (inverse du gonflement, qui libérerait des fractures ouvrant la voie à la remontée du gaz vers la surface) [24]. Nous avons ensuite étudié la structure et la dynamique interfoliaire pour les différents états métastables. Dans les pores de plus grande taille (4 -10 nm), nous avons montré la coexistence de deux phases (riche ou pauvre en $CO_2$), comme dans le réservoir (bulle de $CO_2$ au-dessus d'un aquifère), mais les surfaces restent toujours hydratées à cause de la présence des cations compensateurs.

*Echange cationique*

L'un des enjeux pour prédire le transport et la rétention des ions est la compréhension de l'échange ionique, c'est-à-dire le remplacement d'un cation interfoliaire par un cation en solution (FIGURE 7). Dans le cas de l'échange de $Na^+$ par $Cs^+$, la réaction est thermodynamiquement favorable et exothermique. La thermodynamique de l'échange ionique peut être comprise en analysant les rôles respectifs de la phase aqueuse et de la phase interfoliaire. Par simulation moléculaire, comparée à des expériences de microcalorimétrie, nous avons montré que la force motrice de cet échange n'est pas à trouver dans des interactions préférentielles entre le césium et les surfaces d'argile, mais dans la plus grande enthalpie d'hydratation du sodium (comparé à $Cs^+$) qui est relâché dans la solution [25]. Par exemple, pour une montmorillonite sodique avec six molécules d'eau par cation, l'échange d'un cation $Na^+$ par un $Cs^+$ entre la phase argile et la phase gaz, qui sert de référence (voir FIGURE 7), est associé à un coût d'enthalpie libre de +120 kJ/mol, et est donc très défavorable. C'est le remplacement du $Cs^+$ par $Na^+$ dans la phase aqueuse (-124 kJ/mol pour l'échange avec la phase gaz) qui rend l'échange entre l'argile et la solution thermodynamiquement favorable (bilan de la réaction: $\Delta_r G$ = -4 kJ/mol). Nous avons par ailleurs proposé la première étude au niveau moléculaire du processus d'échange ionique, ce qui a nécessité d'inclure explicitement les bords de feuillets dans nos simulations [26].

*Hydrophilie / hydrophobie du talc*

Les propriétés de mouillage des sols et des roches jouent un rôle crucial dans le transport, et donc la disponibilité, de l'eau et du pétrole. Nous avons caractérisé par simulation moléculaire le caractère hydrophile/hydrophobe des surfaces d'argile électriquement neutres telles que le talc. Le talc se comporte expérimentalement comme hydrophobe (angles de contact élevés, de l'ordre de 85° sur des monocristaux macroscopiques) mais présente également des sites très hydrophiles (adsorption de vapeur d'eau, avec une enthalpie libre d'adsorption d'environ 30 kJ/mol, cohérent avec la formation d'une liaison hydrogène entre un groupe hydroxyle de surface et la molécule d'eau) [27]. Par diverses analyses, telles que l'étude des fluctuations extrêmes de la densité au voisinage des surfaces, nous avons montré qu'il n'y avait pas de contradiction entre ces deux comportements, en précisant la compétition entre les interactions eau/surface et eau/eau. Alors que la compétition entre l'adsorption à la surface et l'entropie gagnée à rester dans la phase gazeuse détermine le taux d'occupation des sites à faible humidité relative, c'est la compétition entre l'adhésion (interaction eau-surface) et la cohésion (eau-eau) qui détermine le comportement hydrophile ou hydrophobe à saturation (humidité relative de 100%, où coexistent l'eau liquide et sa vapeur) [28].

**Conclusion et perspectives**

A l'instar des expérimentateurs, simulateurs et théoriciens doivent adapter leurs concepts et outils en fonction de l'échelle d'observation considérée - et bien sûr de la question à laquelle ils essaient de répondre. La modélisation multi-échelle (non seulement à plusieurs échelles mais aussi en faisant le lien entre elles) permet ainsi d'aborder de nombreux problèmes de physico-chimie. Ce type d'approche est particulièrement utile pour l'étude des milieux poreux chargés tels que les argiles, dont les propriétés découlent de phénomènes à des échelles d'espace et de temps couvrant plusieurs ordres de grandeur. Nous appliquons également cette stratégie pour les liquides ioniques dans les carbones nanoporeux, afin de faire le lien entre les travaux menés ces dernières années au laboratoire à l'échelle moléculaire et les résultats expérimentaux à l'échelle macroscopique (impédance électrochimique et micro-balance à quartz notamment). D'un point de vue plus fondamental, si les outils pour décrire les effets spécifiques à l'échelle moléculaire sont relativement bien établis, la diversité des phénomènes qui se manifestent à l'échelle mésoscopique nécessite le développement d'outils et de concepts spécifiques, dont le développement plus récent explique peut-être la grande variété des approches développées à cette échelle [29]. Dans ce domaine, il serait donc souhaitable dans les années qui viennent de comparer les mérites et limitations de ces différentes méthodes, en particulier dans la perspective d'une démarche multi-échelle globale.

**Bibliographie**

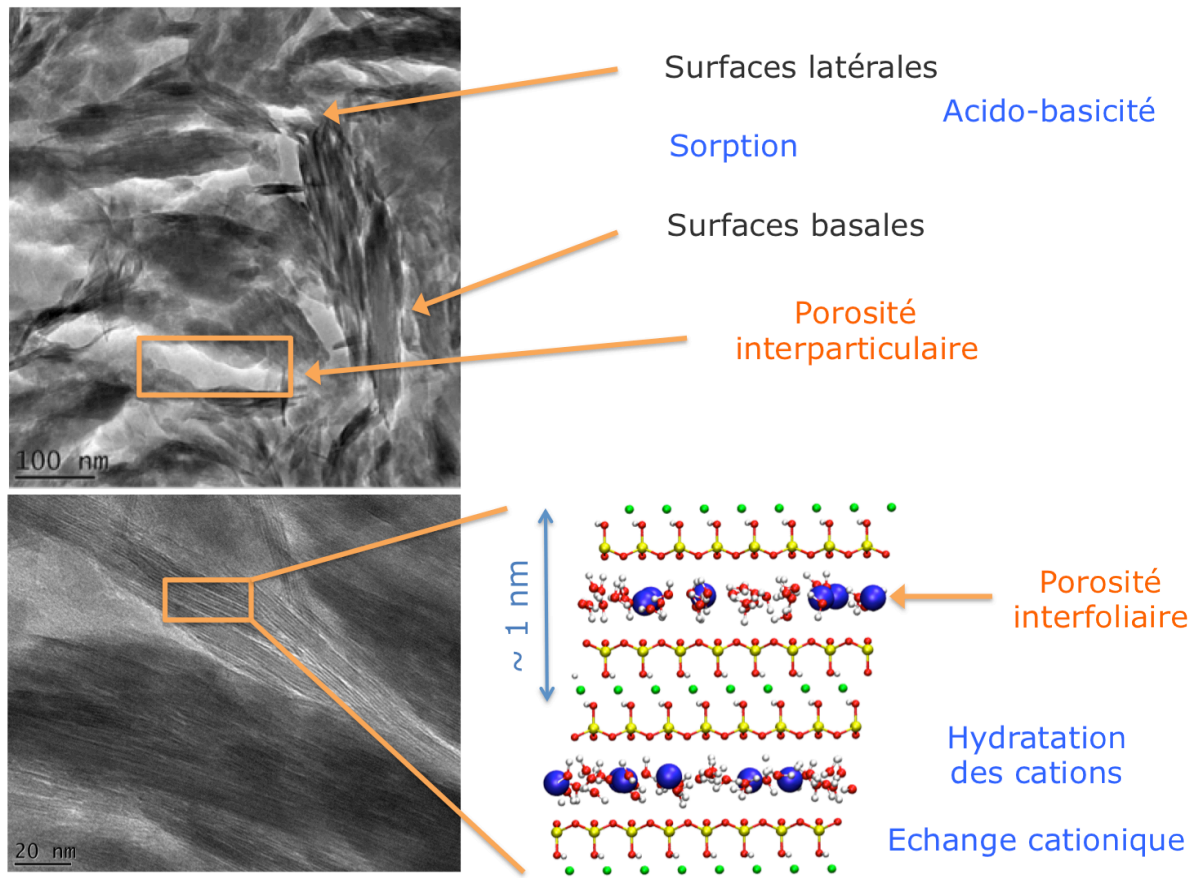

**FIGURE 1** – Structure multi-échelle des argiles : l'empilement de feuillets d'aluminosilicates forme des particules présentant des surfaces externes basales et latérales aux propriétés physico-chimiques distinctes et une nanoporosité interfoliaire, qui contient, comme les autres porosités, de l'eau et des cations (Na$^+$ sur le schéma de droite). Les images de Microscopie Electronique en Transmission à Haute Résolution d'une Illite du Puy ont été réalisées par Stéphane Gaboreau (BRGM) dans le cadre du projet ANR SIMISOL.

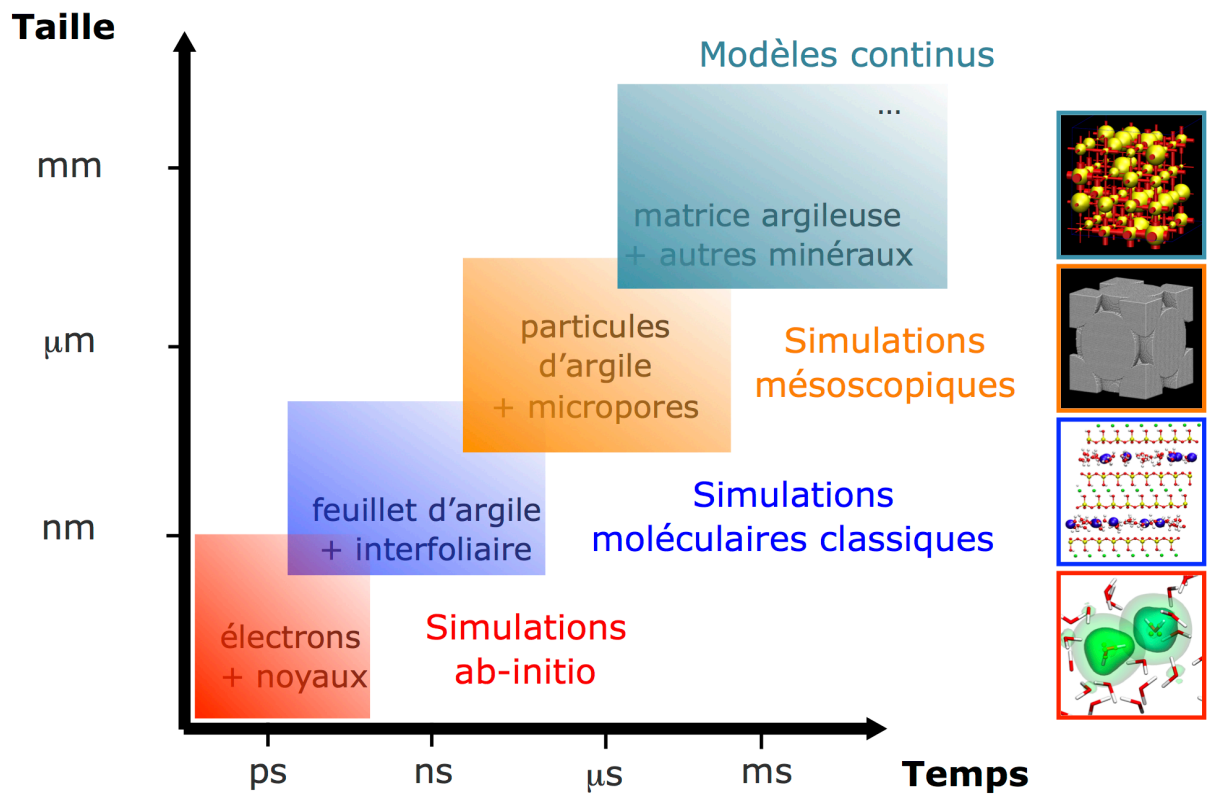

**FIGURE 2** – Démarche de modélisation multi-échelle, illustrée dans le cas des argiles. Selon les échelles de temps et d'espace pertinentes pour répondre à une question de physico-chimie donnée, on adoptera le niveau de description adaptée. Il faut de plus faire le lien entre les différents niveaux de modélisation. On peut ainsi exploiter les descriptions aux échelles les plus fines pour obtenir des modèles plus simples mais qui restent pertinents aux échelles supérieures.

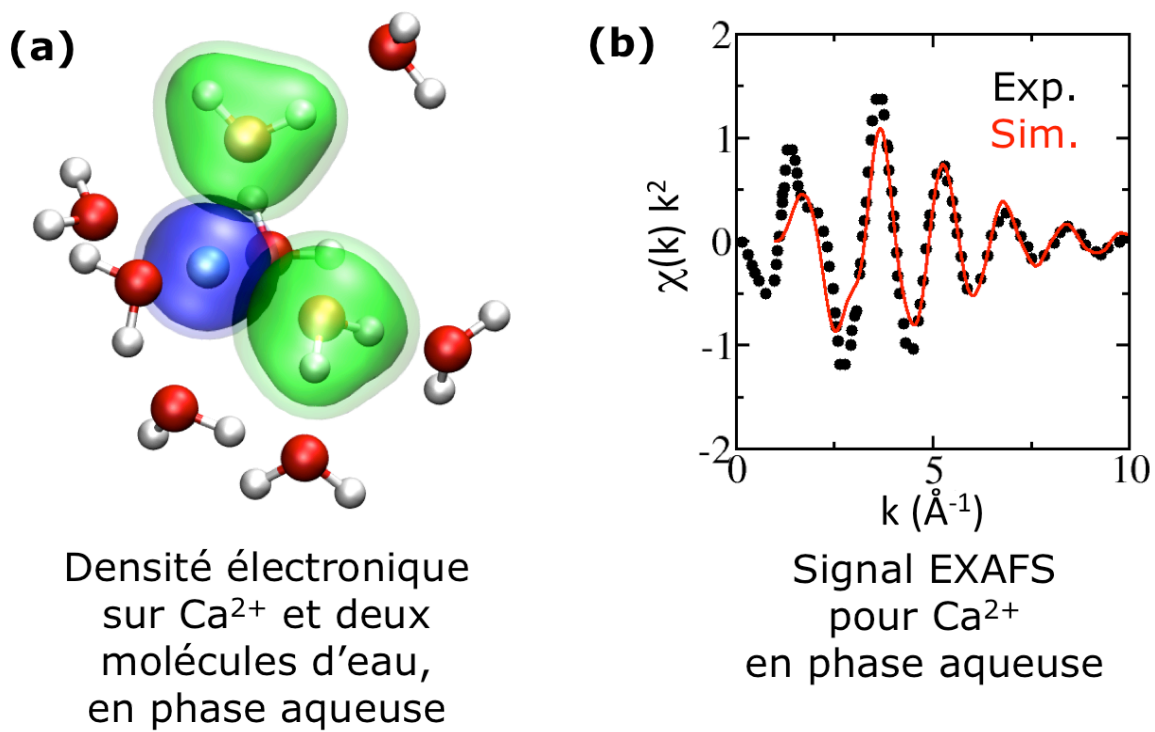

**FIGURE 3** – (a) A partir de calculs *ab initio*, il est possible de déterminer la densité électronique associée à chaque ion ou molécule en phase condensée (ici un ion calcium et deux molécules d'eau, en phase aqueuse), ainsi que la force agissant sur chaque noyau. A partir de ces informations, il est possible de paramétrer des champs de forces classiques de très bonne qualité. (b) A partir d'une trajectoire de dynamique moléculaire classique obtenue avec un tel champ de force (qu'il serait impossible d'obtenir directement par dynamique moléculaire *ab initio*), on peut calculer les propriétés du système. Ici, le signal EXAFS (*) pour l'ion ion $Ca^{2+}$ dans l'eau est en très bon accord avec les résultats expérimentaux, tant pour la période des oscillations qui traduit la distance de l'ion aux molécules dans sa sphère d'hydratation, que pour l'amplitude, qui reflète le nombre de molécules voisines. D'après [11].

(*) On a tracé ici, comme cela se fait habituellement pour la spectroscopie EXAFS (Extended X-ray Absorption Fine Structure), l'absorption normalisée $\chi(k)$ multipliée par le carré du vecteur d'onde k.

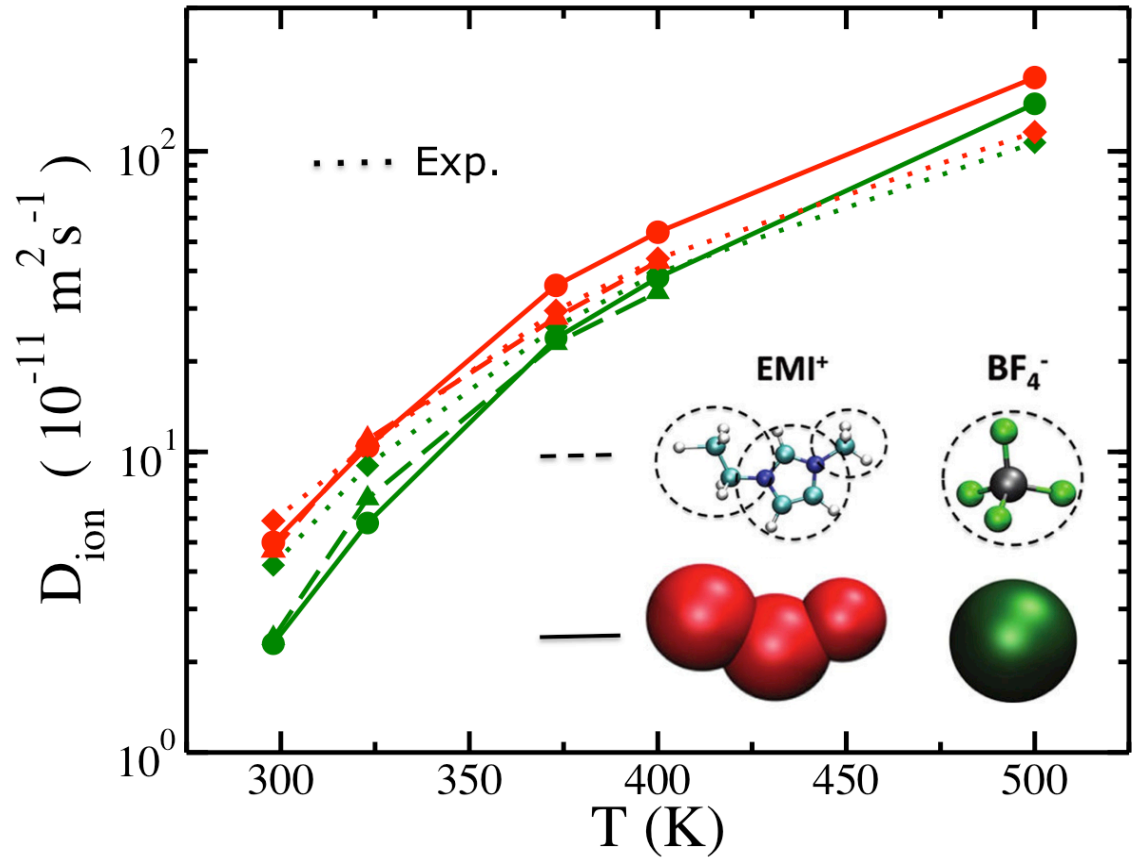

**FIGURE 4** – Pour simplifier le calcul des interactions entre deux molécules, on peut représenter celles-ci par un petit nombre de sites (ici 3 pour les cations et 1 pour les anions) qui interagissent par des potentiels effectifs, ou "gros grains", au lieu de représenter explicitement tous les atomes. Un modèle développé ici (courbes en trait plein) pour le tétrafluoroborate ($BF_4^-$) de 1-éthyl-3-méthylimidazolium ($EMI^+$), un liquide ionique étudié pour son utilisation dans les supercondensateurs à base de carbone nanoporeux, reproduit de nombreuses propriétés expérimentales, ici le coefficient de diffusion (courbes en pointillé) des cations en rouge, et des anions, en vert, en fonction de la température, aussi bien que le modèle "tout-atome" (courbes en tirets) dont il est issu. D'après [14].

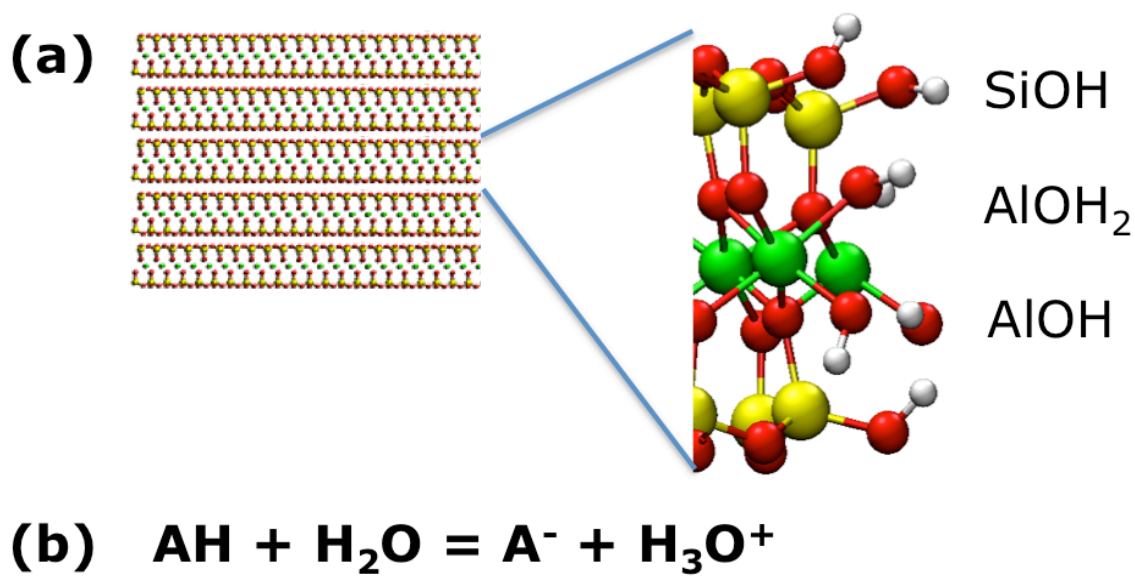

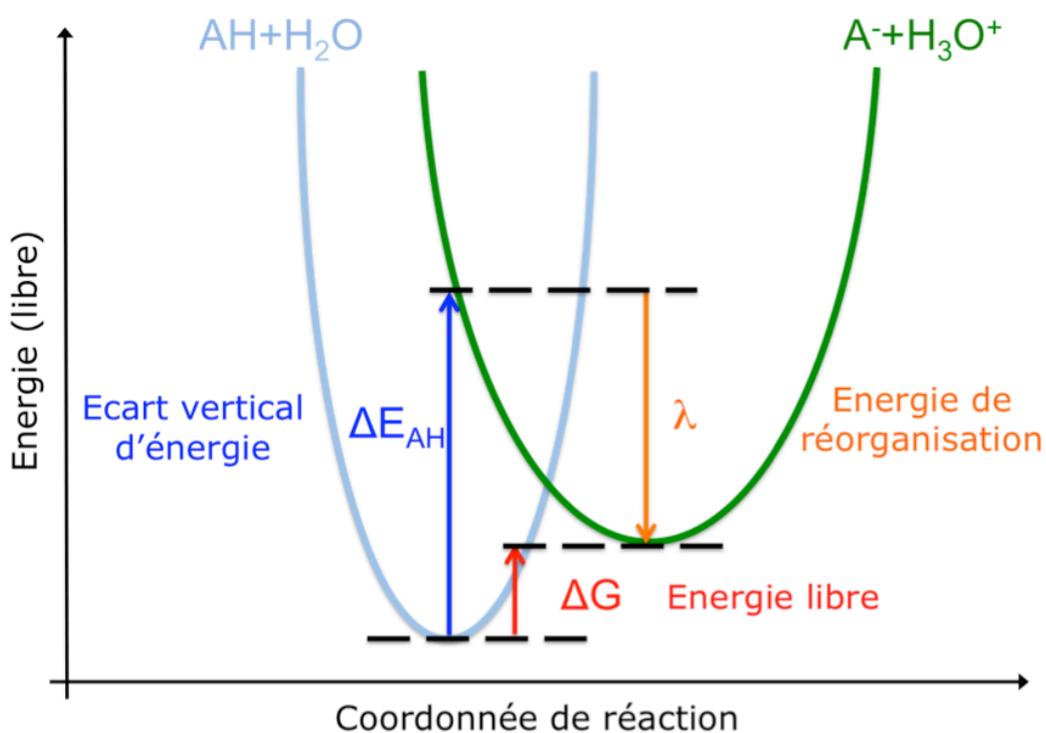

**FIGURE 5** – (a) Les surfaces latérales des particules d'argiles présentent des sites (ici des sites aluminols et silanols sur la face 010 de la pyrophyllite) aux propriétés acido-basiques. Leur état de protonation en fonction du pH joue un rôle essentiel sur la rétention des cations sur ces sites (aux conséquences importantes dans le contexte du piégeage de métaux lourds ou d'éléments radioactifs). (b) La théorie de Marcus du transfert électronique peut être adaptée à la réaction de transfert de proton. On peut alors distinguer l'énergie verticale de déprotonation

ΔE$_{AH}$, qui correspond à l'énergie nécessaire pour transférer le proton dans la phase aqueuse sans réorganiser le site après la déprotonation, de l'énergie de réorganisation λ, à laquelle contribue tant les modifications de longueur des liaisons au voisinage du site que les réorganisation du solvant après déprotonation. L'énergie libre de la réaction, directement liée au pK$_a$ du site correspondant, est la somme de ces deux contributions. D'après [23]. La cinétique du processus pourrait également être étudiée en examinant l'enthalpie libre d'activation, c'est-à-dire la différence entre d'une part l'intersection entre les courbes bleue et verte, correspondant respectivement aux réactifs et aux produits de la réaction, et d'autre part le minimum pour les réactifs.

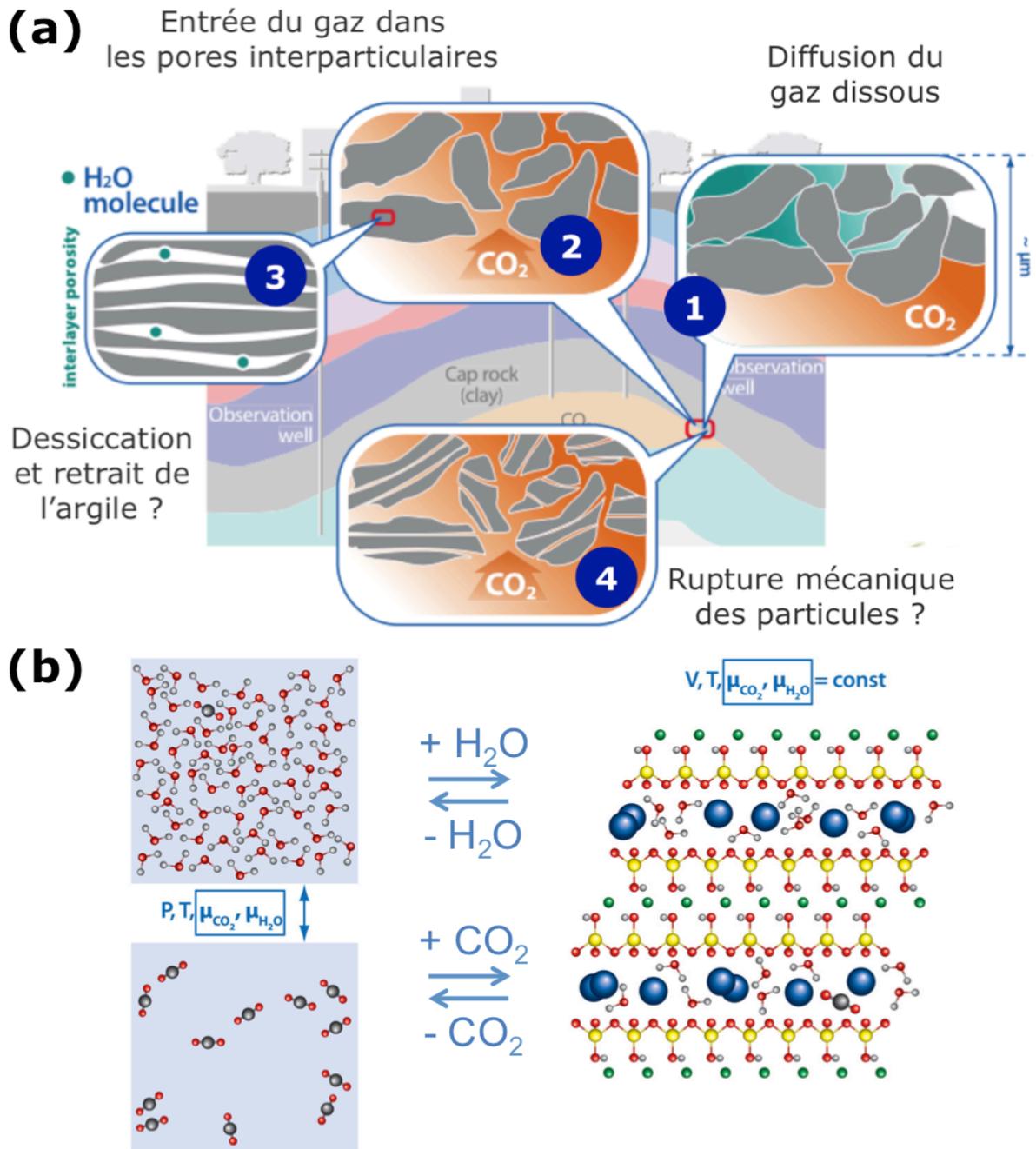

**FIGURE 6** – (a) Différentes options sont à l'étude pour la séquestration du $CO_2$ dans le sous-sol, afin de limiter les émissions de ce gaz à effet de serre dans l'atmosphère. L'une d'entre elle est de l'injecter sous forme supercritique dans des aquifères salins (donc impropres à la consommation). Le fluide supercritique présente une densité bien plus élevée que le gaz, et il est donc possible de stocker sous cette forme une plus grande quantité de $CO_2$. Le $CO_2$ se trouverait alors sous forme dissoute dans l'eau présente dans la formation rocheuse, et dans

une deuxième phase, moins dense que l'eau et plus riche en $CO_2$, qui remonterait vers la surface en l'absence d'une roche couverture imperméable. Différents phénomènes sont à considérer pour le transport du $CO_2$ à travers cette couverture, généralement argileuse, lors de la mise en contact avec le réservoir situé en-dessous. (b) Les simulations moléculaires dans l'ensemble grand-canonique permettent de déterminer la composition du fluide interfoliaire et le gonflement ou retrait éventuel de l'argile, en fonction des conditions du réservoir - où deux phases fluides (riche en eau et riche en $CO_2$) sont en équilibre à une température et une pression qui dépendent de la profondeur, fixant ainsi les potentiels chimiques des deux espèces.

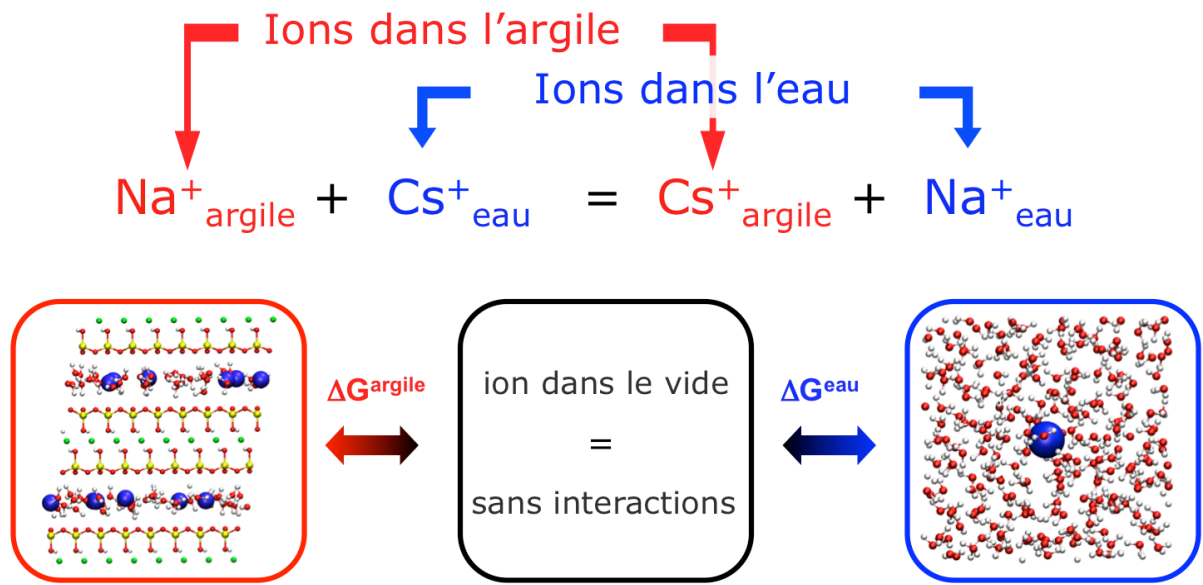

**FIGURE 7** – L'échange ionique consiste à remplacer un ion présent dans l'argile par un autre en solution (ici $Na^+$, contre-ion naturel de l'argile est remplacé par $Cs^+$, radionucléide potentiel). La thermodynamique de cet échange peut être comprise en comparant les contributions relatives du remplacement d'un ion par l'autre d'une part dans l'argile, d'autre part dans la phase aqueuse. Ces deux contributions peuvent être calculées directement par simulation moléculaire, en utilisant la méthode d'intégration thermodynamique, ou déterminées de façon indirecte expérimentalement, en utilisant un cycle thermodynamique.

**Encadré 1 : Simulation ab initio et simulation moléculaire classique**

La dynamique moléculaire *ab initio*, qui nécessite le calcul à chaque pas de temps de la simulation de la densité électronique autour des noyaux, est à ce jour la méthode la plus précise pour le calcul des interactions dans ce type d'approche. Elle repose sur la théorie de la fonctionnelle de la densité, DFT, (on ne considèrera pas ici les approches plus fines de la chimie quantique, qui pour l'instant ne peuvent en pratique pas être appliquées à des systèmes de grande taille et pour de nombreuses configurations) et permet d'étudier des systèmes de quelques centaines d'atomes pendant quelques dizaines ou centaines de picosecondes. Elle permet de plus de prendre en compte la rupture et la formation de liaisons. Pour les systèmes de plus grande taille et les durées plus longues, il est nécessaire de recourir à des descriptions simplifiées, où les degrés de liberté électroniques sont pris en compte dans un "champ de force", caractérisé d'une part par une forme analytique (comme les potentiels de Coulomb ou Lennard-Jones) et d'autre part par des jeux de paramètres associés aux différentes espèces (par exemple la charge partielle ou le diamètre de Lennard-Jones). L'analyse de ces simulations avec les outils de la physique statistique permet de remonter ensuite aux propriétés du système. Pour plus de détails, on pourra se reporter au numéro spécial de l'Actualité Chimique consacré à la modélisation, en particulier [A] pour la simulation moléculaire en général, et [B] pour un exemple d'application aux liquides ioniques.

[A] Borgis D., Boutin A., Vuilleumier R. NUMERO SPECIAL ACTUALITE CHIMIQUE A COMPLETER

[B] Padua A.A.H. NUMERO SPECIAL ACTUALITE CHIMIQUE A COMPLETER

**Encadré 2 : Simulations mésoscopiques sur réseau**

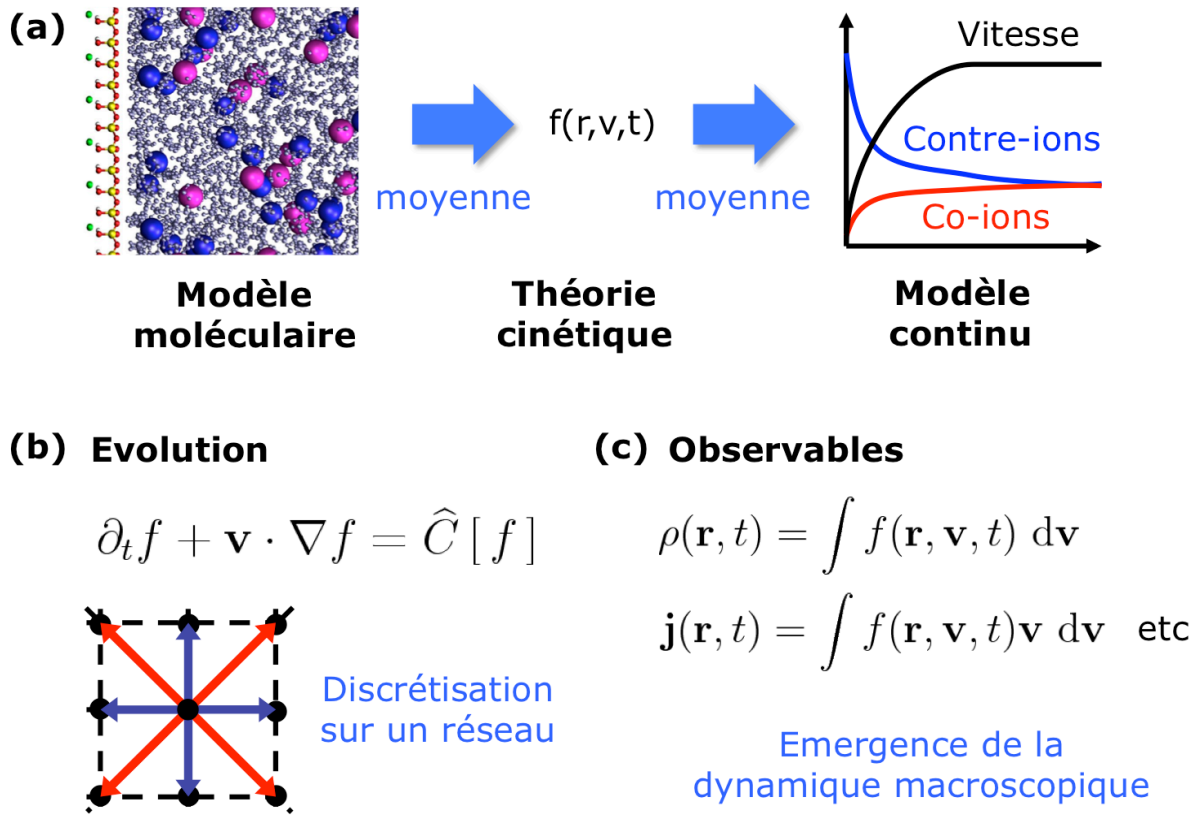

FIGURE ENCADRE 2 – (a) Entre une description moléculaire où la trajectoire de chaque atome est suivie au cours du temps et une description continue en termes de champs (densités, vitesse, ...) se trouve la théorie cinétique, dont l'élément central est la densité de probabilité f($\mathbf{r}$,$\mathbf{v}$,t) de particules à une position $\mathbf{r}$ avec une vitesse $\mathbf{v}$ à un instant donné t. (b) Cette fonction évolue selon une équation dite cinétique, dont le membre de gauche décrit l'effet du transport, et dont le membre de droite, appelé "opérateur de collision", traduit l'effet de toutes les interactions au sein du système. En pratique, cette équation est discrétisée: on interprète alors le système comme composé des particules évoluant sur les nœuds d'un réseau, et ne pouvant avoir qu'un nombre fini de vitesses possibles, qui correspondent au passage d'un nœud à l'un de ses voisins directs. Historiquement, de tels "automates cellulaires" obéissant à de telles règles sont d'ailleurs apparus avant les versions plus modernes, de type dit de "Boltzmann sur réseau" (Lattice Boltzmann), obtenues en discrétisant une équation cinétique.

(c) Les propriétés observables telles que la densité $\rho(\mathbf{r},t)$ ou le flux $\mathbf{j}(\mathbf{r},t)$ de particules à une position et un instant donnés sont donnés par les moments de f($\mathbf{r},\mathbf{v}$,t) par rapport aux vitesses, soit des intégrales dans le cas continu, soit des sommes simples dans le cas discret. L'évolution de ces observables macroscopiques émergent de l'évolution de f($\mathbf{r},\mathbf{v}$,t). Ainsi, dans le cas le plus simple (où l'opérateur de collision fait relaxer f vers la solution d'équilibre local de Maxwell-Boltzmann), la densité, le flux et le tenseur des contraintes vérifient l'équation de Navier-Stokes. L'algorithme Lattice Boltzmann est ainsi couramment utilisé pour simuler les écoulements dans des géométries ou des régimes hydrodynamiques complexes.